\documentclass[12pt]{amsart}
\usepackage{hyperref} 
\usepackage[utf8]{inputenc}
\usepackage[russian,english]{babel}
\author{Alexander Givental}
\title{On Quantum Measurement}
\def\h{\hbar}
\def\RR{\mathbb R}
\def\PP{\mathbb P}
\def\H{\mathcal H}
\def\bra{\langle}
\def\ket{\rangle}
\def\tr{\operatorname{tr}}
\begin{document}

\begin{abstract} We propose a solution to the quantum measurement paradox by
first identifying its classical counterpart.
\end{abstract}

\maketitle

In the standard after John von Neumann \cite{vonN} description of quantum mechanical systems the continuous, deterministic, and time-reversible evolution of a wave function governed by the Schr\"odinger equation is interrupted by discrete and irreversible ``collapses'' of the function which are caused by the acts of measurement committed by a classical macroscopic observer. This description is pretty much at odds with the paradigm of classical mechanics, where Hamilton equations govern the deterministic evolution of points in the phase space, while the observer is completely excluded from the picture. The discrepancy is viewed as a paradox leading to the ``measurement problem'': how to reconcile the collapse with the fact that the observer as a physical object is in principle also subject to Schr\"odinger's evolution? After almost a century of debates and developments (see \cite{Arch}) the fate of this issue begins to dangerously resemble the pathetic lot of its philosophical companions (the ``free will'' and alike),
because for a foundational question like this, one is entitled to expect a prompt and non-technical answer.  

As the myriad of other authors who ever proposed a solution to this problem, we would like to believe that ours is (a) correct and (b) new (of which the former is possible while the latter rather not). Yet, we can at least assure the reader that it is radically simple. Namely, we argue that the same measurement paradox is actually present in classical mechanics, though for historical reasons we are trained not to notice this. The simplicity, however, comes at a cost to our philosophical paradigms, which we try to rethink at the end of this essay. 


{\bf Hamiltonian dynamics vs. statistical mechanics.}
A classical mechanical system is described by its Hamiltonian flow.
This means that the phase manifold $M^{2n}$ of the system carries
symplectic structure $\omega$ allowing one to convert the differential $dH$
of the {\em Hamiltonian} of the system (a.k.a. the total energy function
$H: M\to \RR$) into a vector field $v_H$ whose phase trajectories
(i.e. solutions to the ODE system $\dot{x}=v_H(x)$) define the dynamics.
Alternatively, {\em classical observables} $F$ (i.e. functions $M\to \RR$)
evolve in time according to the equations $\dot{F}=\{ H, F\}$, where $\{ \cdot, \cdot\}$ is the Poisson bracket (defined by the symplectic structure $\omega$ the way explained in symplectic geometry textbooks).

We are writing the equations only to sound mathematical because what we are trying to say needs no formulas: the dynamics is totally deterministic since (due to the uniqueness and existence of solutions of ODE systems) the initial position of the phase point unambiguously determines the point's trajectory, and this description does not involve any observers or acts of measurement. In principle one can imagine the Hamiltonian system of the entire universe which would include all particles of all potential observers (to the detriment to their free will).

The point we are going to make now contains nothing new but is seldom
taken seriously. The above statement of determinism applies to {\bf closed}
systems, and is {\bf conditional}: the phase trajectory is uniquely
determined {\em if} the initial condition is known. But how can the initial condition be known when the system is closed, and in particular does not interact with any measuring device that would allow one to determine the initial phase point? We will discuss later why this idealization --- of a closed system whose initial conditions are nonetheless measured precisely and non-invasively --- is considered acceptable, but for now let us not sweep conceptual difficulties under the rug of practicalism, and admit that in a genuinely closed system the initial condition is unknowable in principle.

This realization immediately leads us into the realm of statistical
mechanics. The symplectic phase space carries {\em Liouville measure} $\Omega:=\omega^{\wedge n}/n!$ which is invariant under any Hamiltonian flow, and ``unknowable'' actually means ``distributed uniformly with respect to $\Omega$''.
However, an observation of the system will ``improve'' the distribution turning it into $\rho \Omega$ (where $\rho$ is some probability density on $M$). Assuming that the act of observation didn't destroy the system, the evolution of the density will proceed under the same Hamiltonian flow according to the equation
$\dot{\rho} = \{\rho, H\}$ until the next act of observation. The latter may change the probability density of finding the system at a given point in the phase space from $\rho$ to some $\tilde{\rho}$, and so on. Thus, the evolution of the ({\em a priori} uniform) distribution in the phase space under the deterministic Hamiltonian flow is interrupted by the acts of observation causing irreversible ``collapses'' of the distribution.

More concretely, suppose that at the moment of measurement
the normalized density in the phase space is $\rho$, and the outcome of the
measurement is that the value of some observable $F$ falls into an interval
$(a,b)$. Then right before the measurement the probability $P$ of the value of $F$ to lie in the interval $(a,b)$ is $\int_M \chi_{(a,b)}(F(x))\rho(x) \Omega$ (here $\chi_{(a,b)}$ is the characteristic function of the interval), and so the collapsed density $\tilde{\rho}(x)=\chi_{(a,b)}(F(x)) \rho(x)/P$.

Most importantly, what the probability density in the phase space signifies is not a physical state of the system, but the observer's knowledge about it, and
the collapse reflects the change of that knowledge in the act of measurement. 


{\bf Classical vs. quantum.} In the formalism of quantum mechanics, a state of a given quantum system is captured by the notorious ``wave function'' $\psi$, which is a non-zero element of a complex Hilbert space $\H$. Yet, multiplying $\psi$ by a phase factor $e^{i \theta}$ does not change the state, and so the phase space of the system is actually the {\em projective space} $\PP (\H)$ of all complex one-dimensional subspaces. Furthermore, quantum observables are represented by self-adjoint operators $F$ on $\H$, and the property of $\psi$ to be an eigenvector of $F$ with the eigenvalue $\lambda$ is interpreted as the latter being the definite value of the physical quantity $F$ at the state $\psi$. The time evolution of states in a {\em closed} system is described by the (linear differential) Schr\"odinger equation whose abstract form is $\sqrt{-1}\h \dot{\psi} = H\psi$. Here $H$ is the {\em Hamilton operator}, the quantum observable analogous to the Hamiltonian in classical mechanics. Alternatively, in the Heisenberg picture, the evolution occurs not in the phase space $\PP(\H)$ but in the space of quantum observables, and is governed by the equation $\sqrt{-1}\h \dot{F}=[H,F]$, where $[\cdot ,\cdot]$ is the commutator. Thus, modulo some technicalities, the evolution (described by the one-parametric group of unitary operators $e^{tH/\sqrt{-1}\h}$) is deterministic, and given the Hamilton operator, the future and the past states of the closed system are determined by its initial state.   

All this is nice an clear, except that so far the states $\psi$ have been disconnected from any physical reality. To make the connection, the {\em Copenhagen interpretation} of quantum mechanics posits that the eigenvalues $\lambda_i$ of an observable $F$ (whose spectrum we assume discrete for the sake of simplicity) are the only possible outcomes of an idealized experiment which measures the value of the physical quantity represented by the operator. Furthermore, in the orthonormal basis $\{ \phi_i\}$ of eigenvectors of $F$, the squared absolute values $|\bra \psi | \phi_i\ket|^2$ of the Fourier coefficients of a normalized ($\bra \psi | \psi \ket =1$) state $\psi$ are the {\em probabilities} of obtaining the outcomes $\lambda_i$ at the state $\psi$. When the measurement spits a particular outcome $\lambda_{i_0}$, the state $\psi$ of the system experiences an immediate {\em collapse} into the new state determined by {\em Born's selection rule}: it becomes $\phi_{i_0}$ in the case when the eigenvalue is simple, and in general the new state is the orthogonal projection (yet to be normalized if one wants to employ it for computing probabilities) of $\psi$ to the eigenspace of $F$ with the eigenvalue $\lambda_{i_0}$. Finally, if the act of measurement didn't destroy the system, further time evolution continues under the Schr\"odinger equation starting with the collapsed state in the role of the initial condition.

{\bf Before looking vs. after.} To emphasize the similarity between the quantum formalism and the classical one, we should introduce {\em mixed} quantum states, which is done in terms of quantum counterparts of distribution densities $\rho$ in classical statistical mechanics --- the so-called {\em density matrices}. A {\em pure} state $\psi$ is represented by the density matrix $|\psi\ket \bra \psi|$, which is the rank 1 orthogonal projector in $\H$ onto the line spanned by $\psi$. For a mixture of pure states $\psi_{\alpha}$ occurring with probabilities $p_{\alpha}$, its density matrix $P:=\sum_{\alpha} p_{\alpha} |\psi_{\alpha}\ket\bra\psi_{\alpha}|$. In general it is a non-negative self-adjoint operator on $\H$ with $\tr P=1$.  The expected value of an observable $F$ in this mixed state is $\tr(FP)$ (which is the counterpart of the classical expected value $\int_M F\rho \Omega$). When an ideal quantum measurement of $F$ is done, but the outcome of it ``has not been looked at yet'' by the observer, the density matrix $P$ turns into the diagonal matrix $\sum_i p_i|\phi_i\ket\bra\phi_i|$ whose eigenvalues $p_i=\bra \phi_i|P|\phi_i\ket$ are the diagonal entries of $P$ in a suitable orthonormal eigenbasis $\{ \phi_i\}$ of $F$. For example, measuring $F$ in a pure state $|\psi\ket \bra\psi|$ results in the ``before-looking'' mixed state $\sum_i |\bra\psi|\phi_i\ket|^2 |\phi_i\ket \bra\phi_i|$, which ``after looking'' collapses into one of the pure states $|\phi_{i_0}\ket\bra\phi_{i_0}|$. In the extreme case (corresponding to the Liouville distribution in classical mechanics) of the scalar density matrix $P=1/N$ (where $N=\dim \H$ is assumed possibly huge but finite), no measurement ``before looking'' would change $P$ since $[P,F]=0$ for all $F$, and only the collapses caused by the observer's looking can change that.      

According to Schur--Horn's inequalities \cite{Sch}, the diagonal of an Hermitian matrix $P$ lies inside the convex hull of all permutations of its eigenvalue array, and is one of such permutations only when $P$ is diagonal. Thus, under ``measurement but before looking'' the convex hull shrinks (unless $[P,F]=0$), implying  that (even before looking) quantum measurements are irreversible. This was already known to von Neumann \cite{vonN} who proved the increase of entropy $-\tr (P\ln P)$ when $[P,F]\neq 0$. It is surprising therefore how many ``paradoxical'' thought experiments (including ``Schr\"odinger's cat'' and ``Wigner's friend'' \cite{Wig}) consider a measured system together with the measuring device as a single quantum meta-system and assume that it can be described by the Schr\"odinger unitary evolution.    

{\bf Wigner vs. von Neumann.}  One of the traditional debates about foundations of quantum mechanics concerns the question whether the collapse is a physical process affecting the actual state of the system (in which case the problem of identifying the process arises) or merely a change of knowledge happening in the observer's consciousness. The above description of collapse leaves room for both. Namely, it is tempting to interpret the ``before looking'' collapse of the density matrix as a physical interaction of the system with the measuring device which takes $P$ into the ``time'' average
\[ \lim_{\Delta \tau \to \infty} \frac{1}{\Delta \tau} \int_{0}^{\Delta \tau} e^{\tau F/\sqrt{-1}\h}\ P\ e^{-\tau F/\sqrt{-1}\h}\ d\tau .\]
Here $\tau$ is the fictitious ``measurement time'' parameter in the process of the Schr\"odinger evolution where the measured observable $F$ plays the role of the Hamilton operator. The point is that the off-diagonal matrix elements $e^{\tau(\lambda_i-\lambda_j)/\sqrt{-1}\h}p_{ij}$ averageate to $0$, and do this quickly provided that the precision $\Delta F=|\lambda_i-\lambda_j|$ of measurement is not excessively high:  $\Delta\tau \Delta F \gg \h \approx 10^{-34} kg\ m^2/ s$.       

On the other hand, the observer's looking at the measurement's outcome changes the state of the observer's knowledge about the system. In the virtual debate between von Neumann, who considered that the moment and locus of the collapse can be placed anywhere on the path from the measuring device to the observer's consciousness, and Eugene Wigner, who once insisted on the latter, we should side with von Neumann in the following sense. As our experience with macroscopic measuring devices (as well as with Schr\"odinger's cats) shows, the outcome of the measurement doesn't change if on its path to our consciousness someone else looks at it before us.        

{\bf Celestial mechanics vs. hydrodynamics.} One important difference between the quantum and classical pictures comes from the fact that, unlike classical observables, quantum observables don't commute. Two consecutive collapses caused by measuring $G$ after measuring $F$ result in an eigenvector of $G$ which has no reason to be an eigenvector of $F$ unless $[F,G]=0$. In the classical world, however, measuring ``without looking'' doesn't change the distribution at all, while ``after looking'' the distribution resulting from measuring $F$ and $G$ is supported on the intersection of the supports resulting from measuring $F$ and $G$ separately, and doesn't depend on the order.

Theoretically speaking, one can use successive classical observations (combined with an accurate description of the phase flow between the moments of observation) in order to narrow down the support of the distribution to a single point in the phase space.  The classical paradigm of a system traveling through the phase space along a well-specified trajectory relies on this methodology, which is considered unproblematic due to its origin in Newtonian celestial mechanics.

The motion of celestial bodies is relatively slow, relatively easy to observe in real time, and virtually impossible to influence. Perhaps the complete integrability of Kepler's two-body problem also helps monitoring more complex configurations.

The situation changes, however, when (even in celestial mechanics) chaos enters the picture in the form of exponential divergence of phase trajectories. The \href{http://web.mit.edu/wisdom/www/hyperion.pdf}{tumbling of Hyperion}, a small egg-shaped moon of Saturn, provides a famous example. It is said that the orientation of Hyperion's axis of spinning is unpredictable for modern computers beyond a 100 day period.

In classical hydrodynamics (where the mathematical model is already infinite dimensional) it is even worse. Will anyone ever measure for me, accurately and non-invasively, the velocity field in the turbulent flow from my kitchen tap (that is, before I decide to turn it off)?

We should recall here Isaac Asimov's principle saying that practical unfeasibility is more fundamental than theoretical possibility, and conclude that beyond the few-body celestial mechanics the situation in the classical world is fundamentally the same as in the quantum one, and not only because of chaos and complexity of the systems, but because of the observer's own interference as well. Even if one believes in the deterministic dynamics of the closed classical universe, one's knowledge of its current state comes from density collapses caused by (often invasive) observations, and remains probabilistic.


{\bf Conservative vs. dissipative.} Probability, as commonly known, is the limit of frequencies when the number of trials tends to infinity. Frequencies of what? Of positive outcomes among all outcomes. What's an outcome? Well, it is a certain event which, having happened, is never going to unhappen. Here is the \href{https://www.youtube.com/watch?v=PanqoHa_B6c}{Hitachi double-slit experiment}: the bright dots, which occur one by one on the screen, are the detected positions of electrons after passing through a magnetic field forcing them into one of two paths. Eventually the dots assemble into the interference pattern prescribed by the wave function. The dots are to stay there until the end of youtube (or the universe, whichever comes first). This ability of ours to register the results of observations in irreversible and hence dissipative acts --- i.e. relying on phenomena usually considered secondary, reducible to conservative fundamental physics --- turns out to be a necessary {\em prerequisite} for doing the latter.

Thus, the image of observer-independent deterministically evolving states of a closed (classical or quantum) system renders the states fundamentally undetectable and hence unpredictable. It should be replaced with the image of conscious observers making irreversible measurements which inevitably alter the state of ({\em the observers' knowledge about}) the system, although between the moments of observation it can be considered evolving deterministically, i.e. remain unchanged modulo the Hamilton or Schr\"dinger evolution, provided that the equations of motion are known and can be accurately solved. 

{\bf QBism vs. Empiriomonism.} The above ``paradign shift'' seems to fit well with the doctrine known as QBism ($=$ {\em Quantum Byessianism}, see the article by Herv\'e Zwirn in \cite{Arch}) at least as far as one aspires to clarify the epistemology of quantum mechanics. However, the description and similarity remain vague until one explains who qualifies in the role of ``conscious observer'', how many are there, and how their individual observations are supposed to correlate.

Real-world ``observers'' teach, write grant proposals, publish papers, attend conferences, lunch with colleagues, etc., and occasionally check their experimental results. The accepted in QBism idealization of this complicated social activity makes individual {\em agents} to use quantum mechanics in order to improve their personal probabilistic expectations about the future, based on their past {\em experiences} and Born's selection rule. This choice (and accusations in ``philosophical solipsism'' which inevitably follow) resembles the late 19th century theory of {\em empirio-criticism} by Ernst Mach and Richard Avenarius, which in 1909 went suddenly under scathing critique in the book {\em Materialism and Empirio-criticism} \cite{Len} by Vladimir Lenin. The actual target of Lenin's attack was his fellow marxist Alexander Bogdanov, who was found guilty of espousing the empiricism of Mach and Avenarius, and could not be exonerated even by the fact that in his \href{http://rumagic.com/ru_zar/sci_philosophy/bogdanov/9/}{empiriomonism},  {\em the objective} emerges from the {\em collectiveness} of human experience. The philosophical viewpoint we outline below is close to Bogdanov's, as on the role of an idealized observer we nominate the civilization entire.

Part of the confusion caused by the unexpected role of subjectivity in physics
comes, in our subjective view, from a dose of mysticism philosophers attach to human consciousness --- something that in the age of ChatGPT shouldn't be hard to dispel. To avoid a debate on whether animals (or subway tourniquets) are conscious and in what sense, let's focus on the aspect of subjectivity which is perhaps the only one relevant to doing physics --- one's inner voice; in Arseny Tarkovsky's \href{https://math.berkeley.edu/~giventh/verse/tarkovsky_echo.pdf}{words}
\begin{quotation}  \selectlanguage{russian} 
\dots привычка \newline Говорить с собою, \newline Спор да перекличка \newline Памяти с судьбою\dots\footnote{\dots the habit of talking to oneself, memory and fate debating and echoing each other\dots}
\end{quotation}
Since inner voice is verbal, it comes not before the ``outer voice'' is enabled. But once the ability to communicate in a language is acquired, one can continue practicing an internal dialogue (or monologue) --- pretty much the same way as after downloading the rules of chess, a neural network can excel in the game by playing {\em ad nauseum} against itself. Thus, human consiousness is essentially a social phenomenon: the learned ability of an individual to carry and nurture a private slice of {\em culture}, the latter being distributed among the individuals in the same sense in which the Internet is distributed among laptops. Therefore, the role of subjectivity in physics amounts, first and foremost, to the role of culture in it.

We have no intention to doubt that our civilization has emerged from (and is a part of) a preexisting universe. Yet, it is also true that all we say about the latter is expressed in concepts of our own making. It is us who decides to distinguish light from darkness or to demote Pluto from the rank of planets. The universe, freed of our asking questions about it, is a featureless ``soup'' undivided into its ingredients by the yins and yangs of our concepts. One can argue that the concepts are not arbitrary and capture objective properties of the external world. And yet, it is the relevance of these properties to us is what makes the concepts introduced. Even the notion of causality merely reflects our interest in predicting the future based on our memory of the past. Though we didn't create Nature, we are at least co-authors of its Laws, which capture not the rules of objectively functioning universe, but rather our collective (and active!) experience of interacting with it.

The following example is to cast doubts on the idea of objectivity of positions occupied by classical mechanical systems in their phase spaces. The angular phase of Mars on its orbit at this very moment is certainly {\em known}, but for most readers of this essay it is distributed uniformly along the circle. It is ``known'' only in the sense that somewhere on Earth there are observatories monitoring it, or at least some experts can calculate it from the earlier recorded initial data. Should no data have been ever recorded, would it still be reasonable to assume that the system {\em is} at a particular phase point (and only our knowledge of it is incomplete)?

George Berkeley, the 18th century predecessor of Mach and Avenarius, considered that in order to {\em exist} things must be {\em observed}. This does not mean he denied the world its objective existence; in the contrary, according to him everything exists because it is observed by God. It seems that the measurement problem in quantum mechanics forces us to rehabilitate Berkeley's anti-marxist views with the only correction that (in the spirit of Bogdanov) {\em God} should be replaced with {\em Culture}.\footnote{I am thankful to Michael Remler and Levi Kitrossky for stimulating discussions and their references to Berkeley and Bogdanov respectively.}

We should ultimately admit that the presence of a universal observer --- our civilization, which actively and purposefully interacts with its environment, is a necessary prerequisite for the universe to acquire any specific features. Physicists' staging elaborate experiments and recording their irreversible outcomes is one of the practices that helps the universal observer to form its subjective reflection (commonly known as ``culture'') of its interaction with the environment. While the models of classical and quantum mechanics serve as convenient idealizations of some of these reflections, the idea of the universal physical system that includes the universal observer is not a legitimate extrapolation.

As a final remark-in-passing, let us note that the universal observer does have a {\em will} --- to actively pursue its unique path, but the question whether the will is {\em free} makes no sense, as make no sense any phase points outside the single phase trajectory of a system whose initial conditions cannot be reset.

%

\enddocument